\documentclass[aps,prl,onecolumn]{revtex4}
\usepackage[latin1]{inputenc}
\usepackage{epsfig}
\usepackage{dcolumn}
\usepackage{bm}
\usepackage{amssymb}

\begin{document}

\title{Phase transitions in optimal search times: how random walkers should combine resetting and flight scales}

\author{Daniel Campos and Vicen\c{c} M\'{e}ndez}
\affiliation{Grup de F\'{\i}sica Estad\'{\i}stica, Departament de F\'{\i}sica. Universitat Aut\`{o}noma de Barcelona, 08193 Bellaterra (Barcelona) Spain }

\begin{abstract}
Recent works have explored the properties of L\'{e}vy flights with resetting in one-dimensional domains and have reported the existence of phase transitions in the phase space of parameters which minimizes the Mean First Passage Time (MFPT) through the origin [Phys. Rev. Lett. 113, 220602 (2014)]. Here we show how actually an interesting dynamics, including also phase transitions for the minimization of the MFPT, can also be obtained without invoking the use of L\'{e}vy statistics but for the simpler case of random walks with exponentially distributed flights of constant speed. We explore this dynamics both in the case of finite and infinite domains, and for different implementations of the resetting mechanism to show that different ways to introduce resetting consistently lead to a quite similar dynamics. The use of exponential flights has the strong advantage that exact solutions can be obtained easily for the MFPT through the origin, so a complete analytical characterization of the system dynamics can be provided. Furthermore, we discuss in detail how the phase transitions observed in random walks with resetting are closely related to several ideas recurrently used in the field of random search theory, in particular to other mechanisms proposed to understand random search in space as mortal random-walks or multi-scale random-walks. As a whole we corroborate that one of the essential ingredients behind MFPT minimization lies in the combination of multiple movement scales (whatever its origin). 
\end{abstract}

\maketitle

Resetting can be defined as the interruption of a process in such a way that it is brought back instantaneously from its present state to another fixed state (presumably the initial one) and allowed to evolve once again from there. In the last years there has been a growing interest for studying classical and well-known stochastic transport processes (e.g. Brownian motion, Continuous-Time Random Walks, etc) when subject to stochastic resetting \cite{evans11,evans12,montero12,kusmierz14,majumdar15}. This idea of allowing the particles to start anew its process after some random time has been claimed to have potential applicability to the study of many different systems. For instance, animal foraging is often constrained by the existence of a homing dynamics that makes individuals return from time to time to its nest \cite{potts12,campos13}, leading to the ecological concept of \textit{central place foraging}. Alternatively, the tendency to revisit well known places in animals can be more complex and be driven by sophisticated memory mechanisms \cite{gautestad11,gautestad13}, a fact which is of particular interest since similar ideas have been used to explain human mobility patterns, too \cite{gonzalez08,song10}. These behavioral mechanisms, driven by higher cognitive processes, require an extension of the concept of resetting whose properties have just began being understood \cite{boyer13,boyer14}. In a different context, the transport of information packets through both wired or wireless networks is often subject to transmission losses (errors, overflow, ...), which may require the transmission of new packets after some time (a mechanism which also coincides with resetting under proper assumptions) for increasing the probability of reaching a target whose location is unknown \cite{gelenbe10}. Similar mechanisms could apply too to beam propagation through optical fiber with random inhomogeneities \cite{montero12} and many other transport processes on graphs and complex media. 

Beyond this practical interest, processes with resetting are attractive from a theoretical perspective as they can induce dramatic changes in the system dynamics if compared to their resetting-absent counterparts. So, resetting facilitates the emergence of nonequilibrium stationary states in different scenarios as in free diffusion \cite{evans14,majumdar15}, in motion subject to potential landscapes \cite{pal15}, or in coagulation-diffusion processes \cite{durang13}. Also, it can induce heavy-tailed decays in the probability distribution of models of transport and/or growth \cite{manrubia99,evans11}. On more mathematical grounds, resetting processes have been found to have interesting connections to some variants of the \textit{multi-armed bandit} problem \cite{dumitriu03,janson12}, which may contribute to extend their potential range of interest.

The focus of the present work is in the use of stochastic resetting as a possible mechanism for optimizing random spatial searches. The relevance of this problem has been already recognized and discussed in some of the aforementioned works (see, e.g., \cite{evans11,evans12,kusmierz14}) but a more specific discussion, addressing the connections of resetting to well known concepts and results from the literature on random search theory, is still lacking. We aim thus at covering this existing gap.

First of all, let us note that search efficiency can be measured in many different ways attending to time, energetic, or other biological or physical constraints \cite{preston10}. In agreement with most statistical physics approaches to the field we identify here efficiency with the time required to reach the target, so we will focus on the Mean First Passage Time (MFPT) of the random walker through the target location as the main magnitude of interest. Previous works have already determined the existence of an optimal value of the resetting rate $r$ which minimizes the MFPT of Brownian walkers through the origin in a semi-infinite media \cite{evans11,evans12}. The extension of this idea to the case of particles moving according to L\'{e}vy Flights has revealed the existence of a first-order transition in the optimal values of the phase space $(r,\mu)$ (with $\mu$ is the L\'{e}vy index) for which the MFPT becomes a minimum. So, for small values of the initial position $x_{0}$ intermediate  optimal values $(r^{*},\mu^{*})=(0.25,1.2893...)$ are found. This regime extends up to a critical value $\left( x_{0} \right)_{\mathrm{cr}}$ such that for $x_{0}> \left( x_{0} \right)_{\mathrm{cr}}$ the optimal combination becomes $(0.22145...,0)$ \cite{kusmierz14}.  Here we will revisit this problem and will show how actually first-order phase transitions (although of a different nature) do also appear also for simpler random-walks, in particular for walks with exponentially distributed flights in finite domains. Furthermore, we will try to establish a connection between these results and other models as mortal random-walks and multi-scale random walks which have been used previously in the random search literature. As a whole, we will show the existence of a common force behind all these motion mechanisms driving MFPT minimization. As we argue here, this force is strongly related to the compromise between efficiently covering nearby regions (\textit{exploitation}) and the \textit{exploration} of new areas. From this perspective, resetting must be viewed as a mechanism that promotes \textit{exploitation} as it allows the possibility to revisit regions which may have been initially missed. This is exactly the same idea behind L\'{e}vy or multi-scale search strategies \cite{viswanathan99,bartumeus14,campos15}. So that, in the forthcoming Sections we will try to understand the reach of this analogy as a way to obtain general principles driving optimization of search processes.

\section{First-passage properties of random-walks with resetting}
We will consider for the moment random walkers moving in a one-dimensional infinite domain (which under proper assumptions provide a convenient representation of many realistic search processes \cite{mendez13,campos15}), and will focus on the role that stochastic resetting plays on the MFPT through the origin, given the walker starts from an arbitrary initial position $x_{0}$. Note in advance that resetting is not expected to be a convenient mechanism for searching unless the target is relatively close to $x_{0}$; otherwise resetting would just prevent the walker from reaching new unexplored areas. So that, the use of an infinite domain with resetting can effectively serve to describe the search for nearby targets in finite domains provided that the domain size is much larger than the typical distance covered before resetting.


In contrast with the cases studied in previous works, we will rather consider here that walkers move according to an isotropic velocity model. So, the flights will not be done instantaneously after some waiting time but the transition occurs progressively at a fixed velocity $+v_{0}$ or $-v_{0}$, each with probability $1/2$. We do so because we consider that a velocity model is probably a more realistic choice for most search applications; anyway, we stress that the extension of our results to the case of instantaneous jumps separated by waiting times would be straightforward.

We will compare the results for two different resetting mechanisms for the sake of completeness. The first one consists of giving the walker the possibility to reset its position to $x_{0}$ whenever a single flight is completed. Accordingly, we will denote by $X_{1},X_{2},...$ the successive positions of the particle after the first, second,... event (where each event can be either a flight or a reset). Thus the position of the particle after the $(i+1)$-th event is chosen according to the rule
\begin{eqnarray}
X_{i+1}= \left\{
\begin{array}{l}
x_{0} ,\quad \textrm{with prob.}\ r\; \textrm{if the i-th event was not a reset} \\
X_{i} \pm v_0 U_{i}, \quad \mathrm{otherwise} \\  
\end{array}\right.
\label{sub}
\end{eqnarray}
Here the flight durations $U_{i}$'s are independent and positive random times determined by the probability distribution function (pdf) $\varphi(t)$, and the symbol $\pm$ denotes that the increment in $X_i$ can be positive or negative with the same probability. In the following, we will term this resetting mechanism as being \textit{subordinated to flights} since the statistics of the flights determines in part the rate at which resetting will occur. Note that this mechanism is slightly different to that proposed in \cite{kusmierz14} since here two consecutive reset events are not allowed, but they must be necessarily separated by at least one flight. This has the advantage that the dynamics in the limit $r \rightarrow 1$ will be nontrivial since there the walker will successively carry out one-flight excursions separated by reset events. Instead, if consecutive resets were allowed as in \cite{kusmierz14}, then for $r \rightarrow 1$ the particle would be kept permanently at $x_{0}$ (which looks somewhat unrealistic) and the MFPT would diverge in that limit.

Alternatively, we will compare this with a second mechanism (which has been already used in \cite{montero12}) in which the statistics of resetting is independent of jumps. In this case flight times are again distributed according to $\varphi(t)$, but the successive times at which a reset occurs are also i.i.d. variables which follow its own pdf $\theta(t)$ (independent from $\varphi(t)$). By introducing this second mechanism (termed as \textit{resetting independent of motion}) we will be able to check if subordination to the motion process has any influence on how resetting affects the values of the MFPT and the dynamics of optimal search.


So that, the evolution of our random walkers will be completely determined for a particular choice of $v_{0}$ and $\varphi(t)$ (also $\theta(t)$, in the case of \textit{motion-independent resetting}). To compute the MFPT we will determine first the rate $q(t;x_{0},x_{0}^{*})$ at which crossings through the origin occur at time $t$ for a free random walk starting from $x_{0}$. The parameter $x_{0}^{*}$ gives the position of the walker immediately after it is reset (for the moment we will set $x_{0}^{*}$ as independent of $x_{0}$ for convenience, while in the end we are mainly interested in the case $x_{0}^{*}=x_{0}$). Then we can apply the renewal property \cite{campos12,campos13}
\begin{equation}
q(t;x_{0},x_{0}^{*})=f(t;x_{0},x_{0}^{*})+\int_{0}^{t} dt' f(t';x_{0},x_{0}^{*}) q(t-t';0,x_{0}^{*})
\label{renewal}
\end{equation}
where $f(t;x_{0})$ is the corresponding first-passage distribution through the origin. So trajectories contributing to $q(t;x_{0},x_{0}^{*})$ are divided into those which cross the origin at $t$ for the first time (first term on the rhs) and those which did it for the first time at a previous time $t'$ (last term on the rhs). The connection of the rate function to the free propagator, for the case of velocity models with constant speed, is simply given by \cite{campos15b}
\begin{eqnarray}
q(t;x_{0},x_{0}^{*})= \left\{
\begin{array}{l}
v_{0} P(0,t;x_{0},x_{0}^{*}), \ \quad \mathrm{if}\ x_{0} \neq 0\\ 
v_{0} P(0,t;x_{0}=0,x_{0}^{*}) - \frac{1}{2} \delta(t), \ \quad \mathrm{if}\ x_{0}=0 \\
\end{array}\right.
\label{rate}
\end{eqnarray}
where the additional term $-\frac{1}{2} \delta(t)$ has been introduced to neglect crossings at $t=0$ by explicitly imposing the condition $q(0;0,x_{0}^{*})=0$.

The MFPT can be obtained easily now by using for instance Laplace transform techniques \cite{campos12,mendez13}. So, if $g(s)$ represents the Laplace transform of an arbitrary time-dependent function $g(t)$, then one obtains from (\ref{renewal}) the expression for the MFPT
\begin{eqnarray}
\langle T \rangle &=& -\lim_{s \rightarrow 0} \frac{d f(s;x_{0},x_{0}^{*})}{ds} = 
-\lim_{s \rightarrow 0} \frac{d}{ds} \left[ \frac{q(s;x_{0},x_{0}^{*})}{1+q(s;x_{0}=0,x_{0}^{*})}\right]\nonumber\\
&=&-\lim_{s \rightarrow 0} \frac{d}{ds} \left[ \frac{v_0P(x=0,s;x_{0},x_{0}^{*})}{1/2 + v_0P(x=0,s;x_{0}=0,x_{0}^{*})} 
\right] 
\label{mfpt}
\end{eqnarray}
which implicitly provides also the expression for the first-passage distribution $f(s;x_{0},x_{0}^{*})=q(s;x_{0},x_{0}^{*})/(1+q(s;0,x_{0}^{*}))$ in the Laplace space.

Note that renewal properties like that in (\ref{renewal}) are strictly valid only for Markovian processes (although they can often be used as a reasonably good approximation for more sophisticated cases, specially regarding MFPT computation \cite{campos12}). So that, in the following we will focus for simplicity on situations which either are Markovian or, equivalently, admit a Markovian embedding.

\subsection{Resetting subordinated to motion}
\label{secsub}
The velocity model driven by the process (\ref{sub}) can be conveniently described through a Continuous-Time Random Walk (CTRW) scheme \cite{montero12,mendez13}. For this we introduce the probability density  $j_{1}(x,t;x_{0},x_{0}^{*})$ for the particles starting a flight from $x$ at time $t$. This allows us to write the mesoscopic balance equation
\begin{eqnarray}
j(x,t;x_{0},x_{0}^{*}) &=& \delta(t) \delta(x-x_{0}) \nonumber\\
&+& (1-r) \int_{0}^{t} dt' \int_{-\infty}^{\infty} dx' \Psi(x',t') j(x-x',t-t';x_{0},x_{0}^{*})\nonumber \\
&+& r \delta(x-x_{0}^{*}) \int_{0}^{t} dt' \int_{-\infty}^{\infty} dx \varphi(t')  j(x,t-t';x_{0},x_{0}^{*})  .
\label{scheme}
\end{eqnarray}
The second and third terms on the rhs of (\ref{scheme}) account for flights starting after another flight or after a reset, respectively. This, together with the initial condition term $\delta(t) \delta(x-x_{0})$ prevents  the particles from being reset until having completed at least its first flight. In the expression (\ref{scheme}) we have introduced  the joint probability $\Psi(x,t)$ of performing a jump of length $x$ with constant speed $v_{0}$ during time $t$. Using that flights are done at speed $v_0$ or $-v_0$, each with the same probability, these pdf's can be easily related to the pdf of flight times through
\begin{eqnarray}
\Psi(x,t)&=&\frac{1}{2}\left[ \delta(x-v_{0}t) + \delta(x+v_{0}t) \right] \varphi(t)\nonumber\\
&=&\frac{1}{2v_0}\delta \left( t- \frac{\vert x \vert}{v_{0}} \right)\varphi(t)
\label{Psi}
\end{eqnarray}


The free propagator for this scheme can be defined as the probability density of particles that are located at point $x$ at time $t$, i.e, 
\begin{equation}
P(x,t;x_{0},x_{0}^{*}) = \int_{0}^{t} dt' \int_{-\infty}^{\infty} dx' \phi(x',t') j(x-x',t-t';x_{0},x_{0}^{*}),
\label{free1}
\end{equation}
where we have defined
\begin{eqnarray}
\phi(x,t) &\equiv& \frac{1}{2} \left[ \delta(x-v_{0}t) + \delta(x+v_{0}t)\right] \int_{t}^{\infty} dt' \varphi(t').
\label{free}
\end{eqnarray}

This last function gives us the probability that a single flight has not finished yet after having travelled during a time $t$ and having covered (either to left or right) a distance $v_{0} t$.
To find an explicit expression for the free propagator we introduce the Fourier-Laplace transform of an arbitrary function $g(x,t)$ as
\begin{equation}
g(k,s)=\int_{0}^{\infty} dt e^{-s t} \int_{-\infty}^{\infty} dx e^{-i k x} g(x,t). 
\end{equation}
In the Laplace-Fourier space the expressions (\ref{scheme}-\ref{free}) become simplified by virtue of the space and time convolution theorems. For example, transforming by Fourier-Laplace Eq. (\ref{scheme}) we obtain
\begin{eqnarray}
\hat{j}(k,s;x_0,x_0^*)&=&e^{-ikx_0}+(1-r)\hat{\Psi}(k,s)\hat{j}(k,s;x_0,x_0^*)\nonumber\\
&+&re^{-ikx_0^*}\hat{j}(k=0,s;x_0,x_0^*)\tilde{\varphi}(s)
\label{tflj}
\end{eqnarray}
where the hat and tilde symbols means Fourier-Laplace and Laplace transforms respectively, and $k$ and $s$ the corresponding Fourier and Laplace arguments respectively.
Setting $k=0$ into Eq. (\ref{tflj}) we can solve for $\hat{j}(k=0,s;x_0,x_0^*)$, which yields
\begin{equation}
\hat{j}(k=0,s;x_0,x_0^*)=\frac{1}{1-\tilde{\varphi}(s)}.
\label{jk0}
\end{equation}
Finally, by transforming Eq. (\ref{free1}) by Fourier-Laplace we obtain $\hat{P}(k,s;x_0,x_0^*)=\hat{\phi}(k,s)\hat{j}(k=0,s;x_0,x_0^*)$, which can be combined with Eq. (\ref{tflj}) to get a closed expression for $P(k,s;x_{0},x_{0}^{*})$ 

\begin{eqnarray}
\hat{P}(k,s;x_{0},x_{0}^{*})=\frac{\hat{\phi}(k,s)}{1-(1-r)\hat{\Psi}(k,s)} \left( e^{-ikx_0}+ \frac{r \tilde{\varphi}(s)}{1-\tilde{\varphi}(s)} e^{-ikx_0^*} \right) ,
\label{Pf}
\end{eqnarray}
where $\hat{\Psi}(k,s)$ and $\hat{\phi}(k,s)$ are given by
\begin{eqnarray}
\hat{\Psi}(k,s)=\frac{1}{2}\left[\tilde{\varphi}(s-ikv_0)+\tilde{\varphi}(s+ikv_0)\right]
\label{Psik}
\end{eqnarray}
and
\begin{eqnarray}
\hat{\phi}(k,s)=\frac{1}{2}\left[\tilde{\varphi}^*(s-ikv_0)+\tilde{\varphi}^*(s+ikv_0)\right].
\label{fi}
\end{eqnarray}

We will focus now for simplicity in the case where the dynamics of flights is Markovian. This corresponds to the case when the flight time distribution is exponential, i.e. 
\begin{equation}
\varphi(t)=\lambda e^{-\lambda  t}.
\end{equation}

Replacing this expression into Eq. (\ref{Pf}) we find after some algebra
\begin{equation}
\hat{P}(k,s;x_{0},x_{0}^{*})=\frac{ (s+ \lambda) e^{-ikx_{0}} +r\lambda \left(1+\frac{\lambda}{s} \right) e^{-ikx_{0}^{*} }} { (s+\lambda) (s+r\lambda) + v_{0}^{2}k^{2}},
\end{equation}
and after inverting the Fourier transform
\begin{eqnarray}
\nonumber \tilde{P}(x,s;x_{0},x_{0}^{*}) &=& \frac{1}{2v_{0}} \sqrt{\frac{s+ \lambda}{s+r \lambda}} e^{-\sqrt{(s+ \lambda)(s+r \lambda)} \vert x-x_{0} \vert /v_{0}} \\
&+& \frac{r \lambda}{2 s v_{0}} \sqrt{ \frac{s+ \lambda}{s+r \lambda}} e^{-\sqrt{(s+ \lambda)(s+r \lambda)} \vert x-x_{0}^{*} \vert /v_{0}} .
\label{prop1}
\end{eqnarray}

Inserting this expression into (\ref{rate}-\ref{mfpt}) we will finally obtain the expression for the first-passage time distribution and the MFPT. For simplicity we just show the result for the case of interest $x_0^{*}=x_0$, which read
\begin{equation}
f(s:x_0,x_0^*)= \frac{s+r\lambda} {r \lambda + s \left( 1+ \sqrt{\frac{s+r \lambda}{s+\lambda}} \right) e^{\sqrt{(s+\lambda) (s+r\lambda)} \vert x_0 \vert /v_0} }
\end{equation}
and
\begin{equation}
\langle T \rangle = \frac{1}{\lambda r} \left( \frac{1+\sqrt{r}} { e^{-\lambda \sqrt{r} \vert x_{0} \vert /v_{0}} } -1 \right).
\label{mfpt1}
\end{equation}

If we compare this last expression with the results obtained in \cite{kusmierz14} for the case of L\'evy flights we observe that the scaling $\lim_{r \rightarrow 0} \langle T \rangle \sim r^{-1/2}$ is kept. However, the universal result for $x_{0} \rightarrow 0$ reported there (see Eq. (8) in Reference \cite{kusmierz14}), i.e. $\lim_{x_{0} \rightarrow 0} \langle T \rangle ( \sqrt{r}-r )^{-1}$, is not recovered. This is a consequence of the rule (\ref{sub}) we have imposed; by forcing the walker to make at least one flight before being reset we have introduced some sort of memory in the system (in the sense that not all events are equivalent) so in that case the Sparre Andersen theorem does not necessarily hold. The different dynamics in both cases is clear by noting that the expression $( \sqrt{r}-r )^{-1}$ leads to the divergence of the MFPT as $r \rightarrow 1$, as we have mentioned above. Instead, it is easy to see from (\ref{mfpt1}) that the MFPT for our case does not diverge in that limit.

\subsection{Resetting independent of motion}
\label{secind}
If the resetting mechanism follows its own time dynamics (according to the time pdf $\theta(t)$, as described above) then it is not convenient to use a Continuous-Time Random Walk scheme as done in the previous section to derive the free propagator of the walk. Instead, we will use the analogy between random walks with resetting and \textit{mortal} random walks. Following the nomenclature of several recent works \cite{yuste13,abad14,campos15b}, \textit{mortal} random walks are defined as walks subject to a mortality function such that after a random time, governed by a given pdf, the particle will disappear (it is, it will 'die'). So, if we define in our random walks with resetting an excursion as the action that goes from $t=0$ to the first reset event or, alternatively, from the $i$-th to the $i+1$-th reset event, then every one of these excursions can be interpreted as an independent mortal random walk. Consequently, the random walk with resetting consists of the time convolution of successive mortal random walks governed by the 'mortality' function $\theta(t)$.

We will focus again for simplicity in the Markovian case, so we choose for the resetting mechanism $\theta(t)=\omega_{m} e^{-\omega_{m} t}$, with $\omega_{m}$ being the frequency of resets (or, alternatively, the 'mortality' rate). In addition we consider 
$\varphi(t)=\lambda e^{-\lambda  t}$, as done in the previous Section.  For this specific choice, the free propagator of a mortal random walk in the Laplace space for an isotropic \textit{velocity model} has been derived before \cite{campos15b} and reads
\begin{equation}
P_{m}(x,s;x_{0})=\frac{1}{2v_{0}} \sqrt{\frac{s+\omega_{m}+\lambda}{s+\omega_{m}}} e^{-\sqrt{(s+\omega_{m}) (s+\omega_{m}+\lambda)} \vert x-x_{0} \vert /v_{0}}
\label{mortal}
\end{equation}
where the subindex $_{m}$ stands for \textit{mortal}.

Using the ideas discussed above and denoting the time convolution operator as $\ast$, we can write the free propagator for the random-walk with resetting as
\begin{eqnarray}
\nonumber P(x,t;x_{0},x_{0}^{*}) &=& P_{m}(x,t;x_{0})+P_{m}(x,t;x_{0}^{*}) \ast \varphi(t) \\
 &+& P_{m}(x,t;x_{0}^{*}) \ast \varphi(t) \ast \varphi(t) + \ldots 
\label{convolution}
\end{eqnarray}
or, transforming to the Laplace space,
\begin{equation}
P(x,s;x_{0},x_{0}^{*})= P_{m}(x,s;x_{0})+ \frac{ P_{m}(x,s;x_{0}^{*}) \varphi(s) } {1- \varphi(s) } .
\label{convlap}
\end{equation}

The first term on the lhs of (\ref{convolution}) accounts for the case where no resetting has occurred yet since $t=0$, the second term stands for the case where only one resetting to the position $x_{0}^{*}$ has occurred up to now, and so on. 

Now, following the same procedure as above, we can insert Equations (\ref{mortal}) and (\ref{convlap}) into (\ref{mfpt}) to provide an expression for the first-passage distribution
\begin{equation}
f(s;x_{0},x_{0}^{*})= \frac{s + \omega_{m}}{\omega_{m} + s \left( 1+ \sqrt{\frac{s+\omega_{m}}{s+\omega_{m}+\lambda}} \right) e^{\sqrt{(s+\omega_{m}) (s+\omega_{m}+\lambda)} \vert x_{0} \vert /v_{0}} } 
\end{equation}
and for its corresponding MFPT
\begin{equation}
\langle T \rangle = \frac{1}{\omega_{m}} \left( \frac{ 1+ \sqrt{\frac{\omega_{m}} {\omega_{m}+\lambda}} } { e^{-\sqrt{\omega_m (\omega_m +\lambda)} \vert x_{0} \vert /v_{0}} } -1 \right).
\label{mfpt2}
\end{equation}

One remarkable point is that one can recover from this result in the diffusive limit ($\lambda \rightarrow \infty$, $v_{0} \rightarrow \infty$ with $v_{0}^{2} / \lambda \rightarrow D$) the expression
\begin{equation}
\langle T \rangle = \frac{1}{\omega_{m}} \left( e^{\sqrt{\omega_{m} /D } \vert x_{0}\vert } -1 \right) 
\end{equation}
which was derived for a Brownian walker with resetting in \cite{evans11}. Likewise, it is easy to check from (\ref{mfpt2}) that $\lim_{\omega_{m} \rightarrow 0} \langle T \rangle= \infty$ and $\lim_{\omega_{m} \rightarrow \infty} \langle T \rangle= \infty$, which guarantees the existence of an optimum resetting rate, except for the case $x_{0}=0$, where the MFPT decays monotonically with $\omega_{m}$ and so the minimum MFPT is obtained for $\omega_{m} \rightarrow \infty$. This qualitatively coincides with the result found for the case of \textit{resetting subordinated to jumps}, which shows that in the limit $x_{0} \rightarrow 0$ the optimal strategy is always resetting as much or as faster as possible. A more complete discussion on this point is provided in the next section.

Again, in the limit $\omega_{m} \rightarrow 0$ we obtain a scaling $\langle T \rangle \sim \omega_{m}^{-1/2}$, which really seems to represent a universal behaviour independent of the mechanism used for resetting. This universal feature is actually reminiscent of the scaling $\sim t^{-1/2}$ for the revisit times to a given position of random walkers which have jump pdf's with finite moments. The first-passage statistics is expected to be asymptotically governed by the probability to be brought back (through a reset) to the initial position after very long excursions. Specifically, MFPT should be inversely proportional to that probability, which leads to $\langle T \rangle \sim \langle t \rangle ^{1/2}$, with $\langle t \rangle$ the mean time for resetting. Consequently, the scaling $\langle T \rangle \sim \omega_{m}^{-1/2}$ follows straightforward.

\section{Optimal search times in infinite domains}
\label{secinf}
Next we will use the results (\ref{mfpt1}) and (\ref{mfpt2}) just derived to analyze the existence of an optimal point in the phase space $(r,\lambda)$ (or $(\omega_{m},\lambda)$) such that it minimizes $\langle T \rangle$ as a function of $x_{0}$. Since we have deliberately chosen a very simple scenario in which both processes (resetting and motion) are characterized by one single scale (together with the spatial scale of the problem, $x_{0}$), one should expect that the optimal strategy is relatively trivial, in comparison with more complex multi-scale or scale-free situations as in L\'{e}vy flights, which surprisingly lead to a first-order transition in the phase space for the optimum $\langle T \rangle$ \cite{kusmierz14}. However, it is still of interest to check whether the different resetting mechanisms proposed will lead to a very different behavior. As we will show in the following Section, the dynamics becomes specially interesting when finite domains are considered (so a new spatial scale is introduced) with the emergence of phase transitions in the phase space $(r,\lambda)$ or $(\omega_{m},\lambda)$.


We start by analyzing the result (\ref{mfpt1}). By differentiating with respect to $\lambda$ one finds that a minimum is found for
\begin{equation}
\lambda_{\mathrm{opt}}= \frac{v_{0}}{\sqrt{r} \vert x_{0} \vert} \left[ 1+ \mathrm{W} \left( \frac{-1}{e (1+\sqrt{r})} \right) \right]
\label{lambdaopt1}
\end{equation}
where $\mathrm{W}()$ denotes the Lambert-W function. One can check that $\lambda_{\mathrm{opt}}$ is always positive for any value in the interval $0<r \leq 1$. Replacing its expression into (\ref{mfpt1}) the global minimum is always found for $r=1$, which corresponds to $\lambda_{opt}= \chi v_{0}/\vert x_{0} \vert$, where we define $\chi \equiv (1+\mathrm{W}(-0.5/e)) \approx 0.768$. So that, the global strategy in this case corresponds trivially to using resetting as much as possible while adapting the flights durations to the spatial scale $x_{0}$.

Next we go for the case of \textit{resetting independent of motion}. Since one can easily check from (\ref{mfpt2}) that the derivative of the MFPT with respect to $\lambda$ is always positive, this immediately leads to the conclusion that the global optimum must be at $\lambda=0$. Differentiation with respect to $\omega_{m}$ then allows one to find that the optimum satisfies $\left( \omega_{m} \right)_{\mathrm{opt}}= \chi v_{0}/\vert x_{0} \vert \approx 0.768 v_{0}/ \vert x_{0} \vert$. It is found then that the optimal resetting rate $\omega_{m}$ coincides with the case of \textit{resetting subordinated to jumps}. This actually shows that the optimal strategy for the random walker in both cases is exactly the same and consists of single-flight excursions interrupted by resetting at the rate $\chi v_{0}/ \vert x_{0} \vert$. These results are actually very intuitive: resetting will be in general a more convenient mechanism for going back to the initial position than turning the direction of motion (at least while resetting is considered to be an instantaneous process), and consequently optimizing the MFPT should rely just on appropriately adjusting resetting to each particular situation. 

We stress that additional mechanisms for resetting have been tested in which the same optimal behaviour has been obtained, again with the same optimal resetting rate $\chi v_{0}/ \vert x_{0} \vert$. So, we can conclude from the above that this is a universal feature for all random walk models with constant velocity $v_{0}$ provided that resetting is Markovian.


\section{Optimal search times in finite domains}
We consider now that the random walkers move in the interval $(0,L)$ with periodic boundary conditions. Note that the procedure to obtain the MFPT described in (\ref{rate}-\ref{mfpt}) is still completely valid, provided that now the free propagator $P(x,t;x_{0},x_{0}^{*})$ is replaced there by the propagator $P_{L}(x,t;x_{0},x_{0}^{*})$ for a finite domain of size $L$. Transforming from one to another is relatively simple just by explicitly imposing the periodic condition at the boundaries \cite{campos15b},
\begin{equation}
P_{L}(x,t;x_{0},x_{0}^{*})= \sum_{m=-\infty}^{\infty} P(x+mL,t;x_{0},x_{0}^{*}),
\label{finite}
\end{equation}
which is valid in the region of interest $x \in (0,L)$. 

Introducing in (\ref{finite}) the expressions for the free propagators we have already derived in previous Sections one can obtain again the corresponding MFPT. Since these calculations do not provide any additional insight into the problem we will just reproduce here the final expressions obtained. For the cases of \textit{subordinated} and \textit{independent} resetting with $x_{0}^{*}=x_{0}$ the MFPT reads, respectively,
\begin{equation}
\langle T \rangle = \frac{1}{\lambda r} \left[ \frac{1+\sqrt{r} + (1-\sqrt{r}) e^{-\lambda \sqrt{r} L/v_0} } { e^{-\lambda \sqrt{r} x_{0}/v_0} + e^{-\lambda \sqrt{r} (L-x_{0})/v_0}} -1 \right]
\label{finite1}
\end{equation}
and
\begin{equation}
\langle T \rangle = \frac{1}{\omega_{m}} \left[\frac{ \left( 1+ \sqrt{\frac{\omega_{m}} {\omega_{m}+\lambda}} \right) \left( 1-e^{-\sqrt{\omega_m (\omega_m +\lambda)} L/v_0} \right)} { e^{-\sqrt{\omega_m (\omega_m +\lambda)} x_{0}/v_0} + e^{-\sqrt{\omega_m (\omega_m +\lambda)} (L-x_{0})/v_0}} -1 \right],
\label{finite2}
\end{equation}
from which (\ref{mfpt1}) and (\ref{mfpt2}) can be recovered, respectively, in the limit $L \rightarrow \infty$.

The main focus of interest now is in understanding how the global optima found in the previous Section for infinite domains changes as the domain size gets reduced. For this, note that in the case where the spatial scales $x_{0}$ and $L$ are of the same order then both resetting and changes of direction will be detrimental, since coming back to $x_0$ is only a good strategy provided there is a target close enough to there, and this will only be the case for $x_{0} \ll L$ (or, equivalently, $\vert L-x_{0} \vert \ll L$). Hence, as long as $L$ decreases and that condition is not fulfilled, we should find that the optimal point in the phase space $(r,\lambda)$, or $(\omega_{m},\lambda)$, will move towards $(0,0)$.

\begin{figure}[h]
\includegraphics{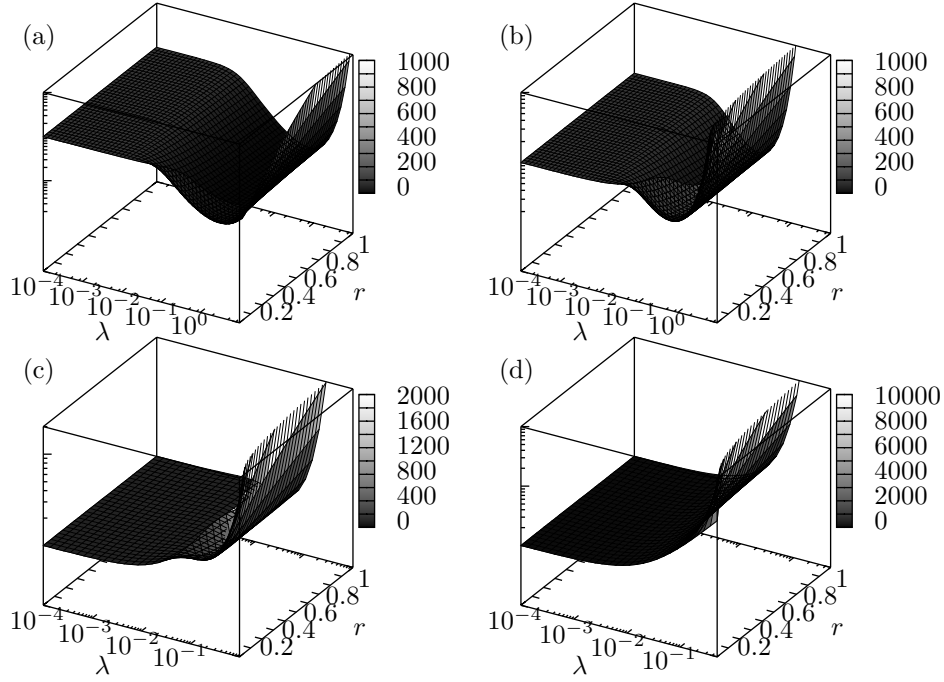}
\caption{MFPT for \textit{resetting subordinated to motion} as a function of the resetting probability $r$ and the average flight length $\lambda$ for different values of the spatial scale: $x_{0}/L=0.005$ (a), $0.025$ (b), $0.1$ (c), $0.25$ (d). The legends on the right of each panel provide information about the order of magnitude of typical values of the MFPT. Arbitrary values $v_{0}=1$ and $L=200$ have been used in all the cases.}
\label{fig1}
\end{figure}

\begin{figure}[h]
\includegraphics{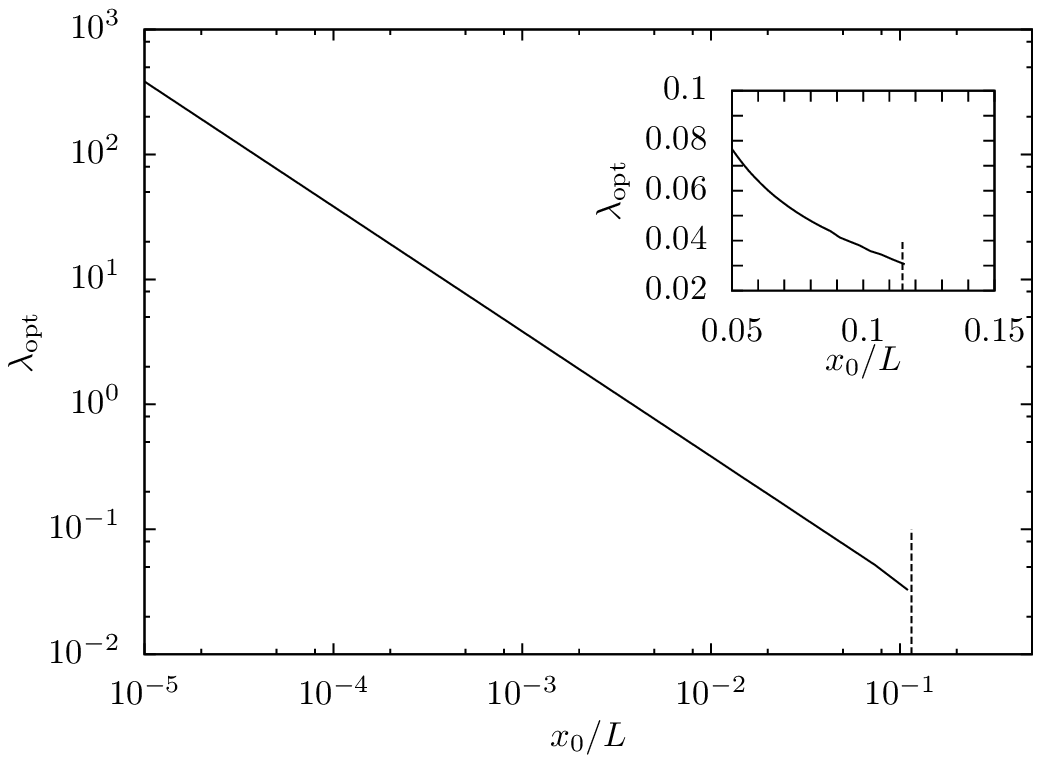}
\caption{Optimal value of the rate $\lambda$ for the case of \textit{resetting subordinated to motion}. The vertical dotted line represents the critical value for which the global optima becomes $(r,\lambda)=(0,0)$. The behaviour close to the critical value is shown in the inset. Arbitrary values $v_{0}=1$ and $L=200$ have been used in all the cases.}
\label{fig2}
\end{figure}

\begin{figure}[h]
\includegraphics{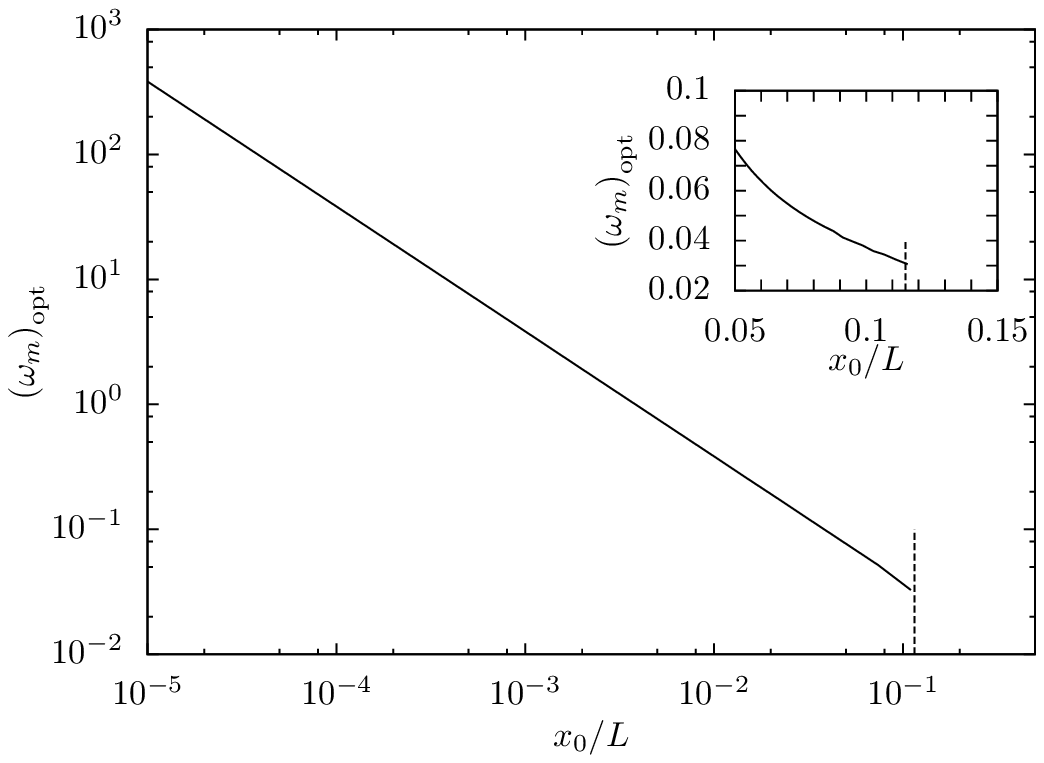}
\caption{Optimal value of the rate $\lambda$ for the case of \textit{resetting independent of motion}. The vertical dotted line represents the critical value for which the global optima becomes $(\omega_{m},\lambda)=(0,0)$. The behaviour close to the critical value is shown in the inset. Arbitrary values $v_{0}=1$ and $L=200$ have been used in all the cases.}
\label{fig3}
\end{figure}

Interestingly, what we obtain from both (\ref{finite1}) and (\ref{finite2}) is that the transition from one regime to the other is not smooth but follows a first-order phase transition. Figure \ref{fig1} provides plots of $\langle T \rangle$ as a function of $r$ and $\lambda$ for \textit{resetting subordinated to motion}. There it can be seen how the point $(0,0)$ represents a global minimum for $x_{0}/L$ large (panel (d)), but when this parameter is small an intermediate minimum appears and eventually it becomes the global minimum (see panels (a) and (b)). This picture can be completed with the information provided in Figures \ref{fig2} and \ref{fig3} (which correspond to the case \textit{subordinated to motion} and \textit{independent of motion}, respectively). There we have computed the optimal value (of $\lambda$ or $\omega_{m}$, respectively) as a function of the order parameter $x_{0}/L$, and we have marked with a vertical dotted line the point at which the phase transition occurs. The scaling $\lambda_{\mathrm{opt}} \sim x_0^{-1}$ and $\left( \omega_{m} \right) _{\mathrm{opt}} \sim x_0^{-1}$ can be also observed there.

The critical point can be easily approximated as follows. First, one should see that the MFPT at the point $(0,0)$ reads $\langle T \rangle =L/2v_{0}$, as can be computed from (\ref{finite1}) or (\ref{finite2}); this coincides with the MFPT one would expect for a ballistic particle moving in the periodic domain $(0,L)$. Then we should determine when the intermediate minimum observed in Figure \ref{fig1}  takes a value equal to $L/2v_{0}$. For doing this, we note that the coordinates of that point can be well approximated by the optimum point we have found in the previous Section (for the case of infinite domains); the justification for this is that we still expect the condition $x_{0}\ll L$ to be valid close to the critical point. Then we will use the points $(r,\lambda)=(1,\chi v_{0}/x_{0})$ for the \textit{subordinated} case, and $(\omega_{m},\lambda)=(\chi v_{0}/x_{0},0)$ for the \textit{independent} case. If we replace this into (\ref{mfpt1}) and (\ref{mfpt2}) we obtain
\begin{equation}
\left( \frac{x_{0}}{L} \right) _{\mathrm{cr}} = \frac{\chi}{2 \left( 2 e^{\chi}-1 \right)} \approx 0.1159..
\label{critical}
\end{equation}
as the critical point at which the transition occurs for both mechanisms. This result coincides well with that reported directly from Eqs. (\ref{finite1},\ref{finite2}) in Figures \ref{fig2} and \ref{fig3} (see insets there). Again, we stress that this result will be valid for any Markovian resetting mechanism provided that random walks are isotropic and the walkers move at a constant speed. 

\subsection{Optimal times with delayed resetting}

We find then that our walkers with constant speed must adopt a very extreme strategy in order to optimize its search times, as they must always interrupt excursions with a reset just after the first flight is completed. One may think that this is simply because resetting is considered here to be an instantaneous process, so it is costless in terms of time. So, we could introduce a delay such that when a reset is carried out the particle must wait an average time $\tau$ before starting a new excursion. This would punish resets so a different (maybe richer) dynamics could be expected.

We have explored this possibility both in the resetting \textit{subordinated to motion} and \textit{independent to motion}, but just show the results of the former for simplicity. In particular, we can replace the last term in Equation (\ref{scheme}) by
\begin{equation}
r \delta(x-x_{0}^{*}) \int_{0}^{t} dt' \int_{-\infty}^{\infty} dx \varphi(t')  j(x,t-t'-\tau;x_{0},x_{0}^{*})  ,
\label{scheme2}
\end{equation}
so we explicitly introduce the delay $\tau$ in the resetting term. With this change now the free propagator in the infinite domain (Eq. (\ref{prop1})) reads
\begin{eqnarray}
\nonumber P(x,s;x_{0},x_{0}^{*}) &=& \frac{1}{2} \sqrt{\frac{s+\lambda}{s+r\lambda}} e^{-\sqrt{(s+\lambda)(s+r\lambda)} \vert x-x_{0} \vert /v_{0}} \\
&+& \frac{r \lambda}{2 \left( s e^{\tau s} + r \lambda (e^{\tau s}-1) \right) } \sqrt{\frac{s+\lambda}{s+r\lambda}} e^{-\sqrt{(s+\lambda)(s+r\lambda)} \vert x-x_{0}^{*} \vert /v_{0}}.
\end{eqnarray}

Using again Eq. (\ref{finite}) and repeating the same procedure as above for obtaining the MFPT, the result reached for $x_0^{*}=x_0$ is
\begin{equation}
\langle T \rangle = (1+ \lambda r \tau) \langle T_{0} \rangle,
\label{delay}
\end{equation}
where $\langle T_{0} \rangle$ is the MFPT in the absence of the delay (given by the expression (\ref{finite1})). Actually, it is possible to extend this result (not shown here) to check that (\ref{delay}) is also valid when the time delay is a random variable that follows any pdf with average $\tau$. 
Likewise, the case of \textit{resetting independent of motion} leads similarly to
\begin{equation}
\langle T \rangle = (1+ \lambda \tau) \langle T_{0} \rangle ,
\label{delay2}
\end{equation}

\begin{figure}[h!]
\includegraphics{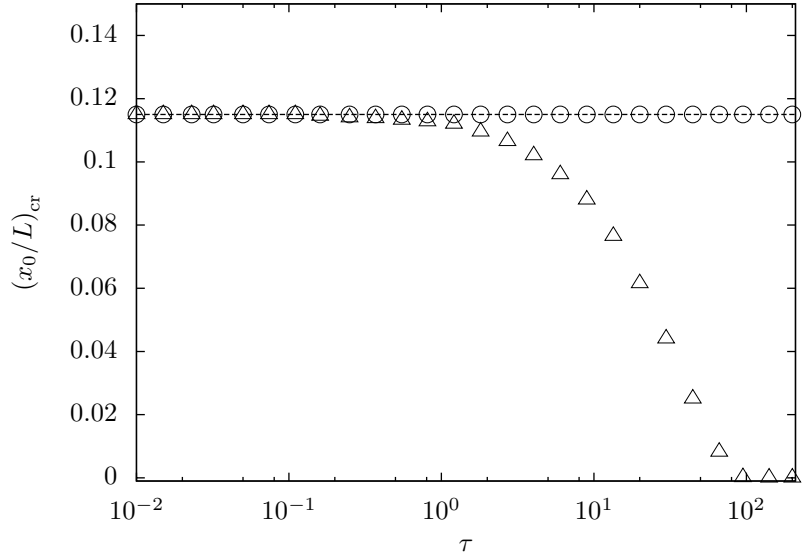}
\caption{Critical value of $x_{0}/L$ as a function of the delay $\tau$ introduced after the reset. Results are shown both for the case of  \textit{resetting subordinated to motion} (triangles) and \textit{resetting independent of motion} (circles) to observe the different dynamics that the delay induces in each case. Arbitrary values $v_{0}=1$ and $L=200$ have been used in all the cases.}
\label{fig4}
\end{figure}

Since the delay only introduces a multiplicative factor in the computation of the MFPT, the effect of this on the optimal strategy is actually minor. One can check (Figure \ref{fig4}, triangles) that the optimal point in the phase space $(r,\lambda)$ for \textit{resetting subordinated to motion} decreases when the delay $\tau$ increases, and eventually for a critical value of $\tau$ the phase transition disappears and the trivial point $(0,0)$ remains as the unique global optima. For the \textit{resetting independent of motion} (Figure \ref{fig4}, circles) the delay does not introduce any change in the dynamics of the optimal MFPT, since the parameter $\omega_{m}$ does not appear in the prefactor in (\ref{delay2}). This reflects that the equivalence between both reset mechanisms studied here break down when additional (non-Markovian) effects as a delay effect are introduced.

\section{Discussion: links between resetting and multi-scale walks}
First-passage time distributions of simple random-walks in infinite media typically decay very slowly following a power-law function or similar, the most famous case being the scaling $\sim t^{3/2}$ found for isotropic Markovian walks, a result known as the Sparre Andersen theorem. Instead, resetting  destroys this scaling by introducing memory effects in the process and leads eventually to an exponential decay of the first-passage pdf, as typically found in finite media (albeit one must be aware that both situations are of a different nature; while the latter reaches an equilibrium state the former asymptotically approaches a nonequilibrium stationary state). As a consequence of this, we have showed here that an optimal resetting rate (in terms of minimizing the MFPT through the origin) does appear. From that point of view, our work extends the case of diffussive movement with resetting which was already explored in references \cite{evans11,evans12} to the case of walkers moving at constant speed $v_{0}$ and carrying out exponentially distributed flights. It is so illustrative to recall the optimum resetting rate found in \cite{evans11}, which in our notation corresponds to $2.538 v_{0}^{2}/\lambda x_{0}^{2}$. Here we have checked that the case of exponential flights leads in all cases to an optimal rate $0.768v_{0}/ \vert x_{0} \vert$. The different scaling found in the two cases comes from the fact that our random walkers follow a velocity model, so characteristic times for the motion process over a distance $x$ are typically $ x /v_0$ (in comparison to Brownian motion where typical times read $x^2/D=\lambda x^2 /v_0^2$).

The case of finite domains we have also explored (now with $x_{0}$ defined as positive) shows that when we introduce the additional spatial scale $L$, then $x_{0}/L$ plays the role of an order parameter such that a first-order phase transition for the optimal resetting rate is found. So, for $x_{0}/L$ small we still can approximate the optimum rate by $0.768v_{0}/x_{0}$, while above the critical point $x_{0}/L \approx 0.1159$ the optimum resetting rate is trivially zero. It is interesting to note that an analogous behavior has been found for the case of two-scale random walks (without resetting) in finite media \cite{campos15}. If the walker is allowed to move according to two characteristic movement scales (say $\lambda_{1}$ and $\lambda_{2}$), then we also find a critical value of $x_{0}/L \approx 0.105$ such that above this value the global optima corresponds to $\lambda_{1}=\lambda_{2}=0$. Instead, for the region under the critical value the best strategy corresponds to $\lambda_{1}=0$ and $\lambda_{2} \approx 0.5v_{0}/x_{0}$. So that, we observe that the role played by this second movement scale $\lambda_{2}$ is similar to the effect that resetting has (although both mechanisms have fundamental differences, since resetting clearly introduces a bias in motion towards the initial point while the two-scale walk is completely isotropic). These results reinforce the idea discussed in previous works on search theory \cite{raposo13,bartumeus14,campos15} that optimizing search efficiency of random walks (in terms of minimizing their MFPT) necessarily implies an appropriate combination of different (at least two) motion scales, which in the limit case leads to the use of free-scale, i.e. L\'{e}vy, strategies. Smaller scales will then be used to efficiently \textit{exploit} closer regions when this contains targets, while larger scales will be used to \textit{explore} further regions. While a more generalized and fundamental study of the link between multiscale motion and MFPT optimization is still elusive, the present work provides so novel and significant evidence in this line.




\textbf{Acknowledgements.} This research has been partially supported
by Grants No. FIS 2012-32334 and SGR 2013-00923. VM also thanks the Isaac Newton Institute for Mathematical Sciences, Cambridge, for support and hospitality during the CGP programme where part of this work was undertaken.

\end{document}